\documentclass[conference]{IEEEtran}
\usepackage{amsfonts}
\usepackage{amssymb}
\usepackage{amsmath}
\usepackage{graphicx}
\usepackage{subfigure}
\usepackage{pst-plot}
\usepackage{pst-node}
\usepackage{pst-fill}
\graphicspath{{figure/}}
\begin{document}
\newtheorem{theorem}{Theorem}
\newtheorem{corollary}{Corollary}
\newtheorem{example}{Example}
\newcommand{\xE}{\textrm{E}}
\newcommand{\xH}{\mathbf{H}}
\newcommand{\xI}{\mathbf{I}}
\newcommand{\xu}{\mathbf{u}}
\newcommand{\xU}{\mathbf{U}}
\newcommand{\xv}{\mathbf{v}}
\newcommand{\xV}{\mathbf{V}}
\newcommand{\xY}{\mathbf{Y}}
\newcommand{\xX}{\mathbf{X}}
\newcommand{\xZ}{\mathbf{Z}}
\long\def\symbolfootnote[#1]#2{\begingroup%
\def\thefootnote{\fnsymbol{footnote}}\footnotetext[#1]{#2}\endgroup}
\title{Feasibility Conditions for Interference Alignment}
\author{\authorblockN{Cenk M. Yetis$^\dagger$}
\authorblockA{Istanbul Technical University\\
Informatics Inst.\\
Maslak, Istanbul, TURKEY\\
Email: cenkmyetis@yahoo.com}
\and
\authorblockN{Syed A. Jafar}
\authorblockA{University of California Irvine\\
Electrical Engineering and Computer Science\\
Irvine, CA, USA\\
Email: syed@uci.edu}
\and
\authorblockN{Ahmet H. Kayran}
\authorblockA{Istanbul Technical University\\
Dept. of Electronics and Comm. Eng.\\
Maslak, Istanbul, TURKEY\\
Email: kayran@itu.edu.tr}}
 \maketitle

\begin{abstract}
The degrees of freedom of MIMO interference networks with constant channel coefficients are not known in general. Determining the feasibility of a linear interference alignment solution is a key step toward solving this open problem. Our approach in this paper is to view the alignment problem as a system of bilinear equations and determine its solvability by comparing the number of equations and the number of variables. To this end, we divide interference alignment problems into two classes - proper and improper. An interference alignment problem is called proper if the number of equations does not exceed the number of variables. Otherwise, it is called improper. Examples are presented to support the intuition that for generic channel matrices, proper systems are almost surely feasible and improper systems are almost surely infeasible.
\end{abstract}
\section{Introduction}
The degrees of freedom (DoF) of wireless interference networks represent the number of interference-free signaling-dimensions in the network. In a network with $K$ transmitters and $K$ receivers and non-degenerate channel conditions, it is well known that $K$ non-interfering spatial signaling dimensions can be created if the transmitters or the receivers are able to jointly process their signals. Cadambe and Jafar \cite{85} recently introduced the idea of interference alignment for the $K$-user wireless interference network with time-varying/frequency-selective channel coefficients, and showed that $K/2$ spatial signaling dimensions are available inspite of the distributed nature of the network which precludes joint processing of signals at transmitters or receivers. While a number of interference alignment solutions have appeared since \cite{85} for different channel settings, many fundamental questions remain unanswered. One such problem is to determine the feasibility of linear interference alignment for MIMO interference networks with constant channel coefficients. It is this open problem that we address in this paper.

\symbolfootnote[2]{The author is supported in part by The Scientific
and Technological Research Council of
      Turkey (TUBITAK). The author is on leave at University of California, Irvine.}

Suppose we denote by $(M\times N, d)^K$, the $K$-user interference network where every transmitter has $M$ antennas, every receiver has $N$ antennas and each user wishes to achieve $d$ DoF. We call such a system a symmetric system. Consider the following examples.

\begin{itemize}
\item $(2\times 2, 1)^3$ - It is shown by \cite{85} that the in the $3$-user interference channel with $2$ antennas at each node, each user can achieve $1$ degree of freedom by linear interference alignment, i.e. by linear beamforming at the transmitters and linear combining at the receivers.

\item $(5\times 5,2)^4$ - Consider the $4$-user interference channel with $5$ antennas at each user. Suppose we wish to achieve $2$ DoF per user, for a total of $8$ network DoF. An analytical solution to this problem is not known but numerical evidence in \cite{55} clearly indicates that a linear interference alignment solution exists. Numerical algorithms are one way to determine the feasibility of linear interference alignment. But, is there a way to analytically predict the feasibility of alignment? In other words, without running the numerical simulation could we have predicted whether a linear interference alignment solution will exist for the $(5\times 5,2)^4$ system?

\item Now consider three distinct systems -$(6\times 4,2)^4, (7\times 3,2)^4, (8\times 2, 2)^4$. Are these systems feasible? Clearly, the last one, $(8\times 2, 2)^4$ is feasible, because simple transmit zero-forcing is enough to eliminate the interference at every receiver. Is the feasibility of $(8\times 2, 2)^4$ case related to the feasibility of other cases where each transmitter successively donates one antenna to the receiver? We will show that these three systems and the $(5\times 5,2)^4$ case belong to the same group.
\end{itemize}

In this paper, we address all these questions, as well as more
complex asymmetric cases where each node may have different number
of antennas and each user may demand different number of degrees of
freedom. The basic approach is to consider the linear interference
alignment problem formulation as a system of bilinear equations. We
determine the correct way to count the number of variables and
equations for a general MIMO interference alignment problem. Then,
based on the number of variables and equations we classify the
system as either proper (number of equations does not exceed number
of variables), or improper (number of equations exceeds the number
of variables), with the intuitive understanding that \emph{proper
systems are almost surely feasible, and improper systems are almost
surely infeasible}.

Aside from the detailed conditions that classify a general (asymmetric) system as proper or improper, our analysis yields a useful rule of thumb for symmetric systems $(M\times N, d)^K$. We find that \emph{donating an antenna from each transmitter to the corresponding receiver, or vice versa, does not change the nature of the system (proper or improper)}, provided every node still has at least $d$ antennas. For example, consider the system $(2\times 3,1)^4$. Is this system proper? While this question may be difficult to answer at first, now suppose we transfer one antenna from each transmitter to its corresponding receiver, to obtain the system $(1\times 4,1)^4$. This system is clearly proper, because simple zero-forcing achieves $1$ degree of freedom for each user in this system. This tells us that the system $(2\times 3,1)^4$ is also proper. Thus, the following are \emph{groups} of proper systems, related by the rule of thumb defined above.

\begin{itemize}
\item $(1\times 4,1)^4, (2\times 3, 1)^4, (3\times 2,1)^4,(4\times 1,1)^4$
\item $(1\times 3,1)^3, (2\times 2, 1)^3, (3\times 1,1)^3$
\item $(8\times 2,2)^4, (7\times 3,2)^4, (6\times 4,2)^4,(5\times 5,2)^4,(4\times 6,2)^4,(3\times 7,2)^4,(2\times 8,2)^4$.
\end{itemize}
In more general terms, the group $(K\times 1,1)^K, ((K-1)\times 2, 1)^K, \cdots, (2\times (K-1),1)^K, (1\times K, 1)^K$ is a proper group. The first and last members of each group are easily seen to be proper (because a  simple zero-forcing solution exists), thereby also determining the status of the rest of the members of the group.  Improper systems can be similarly grouped as well. Completely asymmetric cases require a more sophisticated set of feasibility conditions. The analysis is supported by numerical results for a wide variety of cases, including the specific examples listed above. We begin with the system model.
\section{System Model}
We consider the same $K$-user MIMO interference channel as
considered in \cite{55}. The received signal at the $n^{th}$ channel
use can be written as follows:
\begin{eqnarray*}
\xY^{[k]}(n)=\sum_{l=1}^K\xH^{[kl]}(n)\xX^{[l]}(n)+\xZ^{[k]}(n),
~~
\end{eqnarray*}
$\forall k\in\mathcal{K}\triangleq\{1,2,...,K\}$. Here, $\xY^{[k]}(n) \textrm{ and } \xZ^{[k]}(n)$ are the
$N^{[k]}\times 1$ received signal vector and the zero mean unit
variance circularly symmetric symmetric additive white Gaussian
noise vector (AWGN) at the $k^{th}$ receiver, respectively.
$\xX^{[l]}(n)$ is the $M^{[l]}\times 1$ signal vector transmitted
from the $l^{th}$ transmitter and $\xH^{[kl]}(n)$ is the
$N^{[k]}\times M^{[l]}$ matrix of channel coefficients between the
$l^{th}$ transmitter and the $k^{th}$ receiver.
$\xE[||\xX^{[l]}(n)||^2]=P^{[l]}$ is the transmit power of the
$l^{th}$ transmitter. Hereafter, we omit the channel use index $n$
for the sake of simplicity. The DoF for the $k^{th}$ user's message is denoted by
$d^{[k]}\leq\min(M^{[k]},N^{[k]})$.

\begin{figure}[htb]
\centering
\subfigure[$\left(2\times3,1\right)^2\left(3\times2,1\right)^2$ system in Example \ref{ex:2x3,3x2}.] 
{    \label{fig:Fig1a}
\begin{pspicture}(0,0)(7,6.4) 
\psframe(1,4.8)(1.6,5.8) \pscircle(1.3,5){.1} \pscircle(1.3,5.6){.1}
\pnode(0,5.3){Tx1}
\pnode(1,5){Tx1a}\pnode(1,5.6){Tx1b}\ncline[linestyle=dashed]{Tx1}{Tx1a}\ncline[linestyle=dashed]{Tx1}{Tx1b}
\pnode(1.6,5.3){Tx1c}\psellipse[linestyle=dashed](.5,5.3)(.2,.5)
\rput[l](.1,6){\small{$\xv^{[1]}_1$}}

\psframe(5.4,4.8)(6,5.8) \pscircle(5.7,5){.1} \pscircle(5.7,5.3){.1}
\pscircle(5.7,5.6){.1} \pnode(7,5.3){Rx1}
\pnode(6,5){Rx1a}\pnode(6,5.3){Rx1b}\pnode(6,5.6){Rx1c}\ncline[linestyle=dashed]{Rx1}{Rx1a}\ncline[linestyle=dashed]{Rx1}{Rx1b}\ncline[linestyle=dashed]{Rx1}{Rx1c}
\pnode(5.4,5.3){Rx1d} \psellipse[linestyle=dashed](6.5,5.3)(.2,.5)
\rput[l](6.5,6){\small{$\xu^{[1]}_1$}}

\psframe(1,3.2)(1.6,4.2) \pscircle(1.3,3.4){.1} \pscircle(1.3,4){.1}
\pnode(0,3.7){Tx2}
\pnode(1,3.4){Tx2a}\pnode(1,4){Tx2b}\ncline[linestyle=dashed]{Tx2}{Tx2a}\ncline[linestyle=dashed]{Tx2}{Tx2b}
\pnode(1.6,3.7){Tx2c}\psellipse[linestyle=dashed](.5,3.7)(.2,.5)
\rput[l](.1,4.4){\small{$\xv^{[2]}_1$}}

\psframe(5.4,3.2)(6,4.2) \pscircle(5.7,3.4){.1}
\pscircle(5.7,3.7){.1} \pscircle(5.7,4){.1} \pnode(7,3.7){Rx2}
\pnode(6,3.4){Rx2a}\pnode(6,3.7){Rx2b}\pnode(6,4){Rx2c}\ncline[linestyle=dashed]{Rx2}{Rx2a}\ncline[linestyle=dashed]{Rx2}{Rx2b}\ncline[linestyle=dashed]{Rx2}{Rx2c}
\pnode(5.4,3.7){Rx2d} \psellipse[linestyle=dashed](6.5,3.7)(.2,.5)
\rput[l](6.5,4.4){\small{$\xu^{[2]}_1$}}

\psframe(1,1.6)(1.6,2.6) \pscircle(1.3,1.8){.1}
\pscircle(1.3,2.1){.1} \pscircle(1.3,2.4){.1} \pnode(0,2.1){Tx3}
\pnode(1,1.8){Tx3a}\pnode(1,2.4){Tx3b}\ncline[linestyle=dashed]{Tx3}{Tx3a}\ncline[linestyle=dashed]{Tx3}{Tx3b}
\pnode(1.6,2.1){Tx3c}\psellipse[linestyle=dashed](.5,2.1)(.2,.5)
\rput[l](.1,2.8){\small{$\xv^{[3]}_1$}}

\psframe(5.4,1.6)(6,2.6) \pscircle(5.7,1.8){.1}
\pscircle(5.7,2.4){.1} \pnode(7,2.1){Rx3}
\pnode(6,1.8){Rx3a}\pnode(6,2.1){Rx3b}\pnode(6,2.4){Rx3c}\ncline[linestyle=dashed]{Rx3}{Rx3a}\ncline[linestyle=dashed]{Rx3}{Rx3b}\ncline[linestyle=dashed]{Rx3}{Rx3c}
\pnode(5.4,2.1){Rx3d} \psellipse[linestyle=dashed](6.5,2.1)(.2,.5)
\rput[l](6.5,2.8){\small{$\xu^{[3]}_1$}}

\psframe(1,0)(1.6,1) \pscircle(1.3,.2){.1}\pscircle(1.3,.5){.1}
\pscircle(1.3,.8){.1} \pnode(0,.5){Tx4}
\pnode(1,.2){Tx4a}\pnode(1,.8){Tx4b}\ncline[linestyle=dashed]{Tx4}{Tx4a}\ncline[linestyle=dashed]{Tx4}{Tx4b}
\pnode(1.6,.5){Tx4c}\psellipse[linestyle=dashed](.5,.5)(.2,.5)
\rput[l](.1,1.2){\small{$\xv^{[4]}_1$}}

\psframe(5.4,0)(6,1) \pscircle(5.7,.2){.1} \pscircle(5.7,.8){.1}
\pnode(7,.5){Rx4}
\pnode(6,.2){Rx4a}\pnode(6,.5){Rx4b}\pnode(6,.8){Rx4c}\ncline[linestyle=dashed]{Rx4}{Rx4a}\ncline[linestyle=dashed]{Rx4}{Rx4b}\ncline[linestyle=dashed]{Rx4}{Rx4c}
\pnode(5.4,.5){Rx4d} \psellipse[linestyle=dashed](6.5,.5)(.2,.5)
\rput[l](6.5,1.2){\small{$\xu^{[4]}_1$}}

\ncline{->}{Tx1c}{Rx1d}\ncline{->}{Tx1c}{Rx2d}\ncline{->}{Tx1c}{Rx3d}\ncline{->}{Tx1c}{Rx4d}
\ncline{->}{Tx2c}{Rx2d}\ncline{->}{Tx2c}{Rx1d}\ncline{->}{Tx2c}{Rx3d}\ncline{->}{Tx2c}{Rx4d}
\ncline{->}{Tx3c}{Rx3d}\ncline{->}{Tx3c}{Rx1d}\ncline{->}{Tx3c}{Rx2d}\ncline{->}{Tx3c}{Rx4d}
\ncline{->}{Tx4c}{Rx4d}\ncline{->}{Tx4c}{Rx1d}\ncline{->}{Tx4c}{Rx2d}\ncline{->}{Tx4c}{Rx3d}
\end{pspicture}
}
\centering
\subfigure[$\left(2\times3,1\right)^4$ system in Example
\ref{ex:2x3,1}.] 
{
    \label{fig:Fig1b}
\begin{pspicture}(0,0)(7,6.4) 
\psframe(1,4.8)(1.6,5.8) \pscircle(1.3,5){.1} \pscircle(1.3,5.6){.1}
\pnode(0,5.3){Tx1}
\pnode(1,5){Tx1a}\pnode(1,5.6){Tx1b}\ncline[linestyle=dashed]{Tx1}{Tx1a}\ncline[linestyle=dashed]{Tx1}{Tx1b}
\pnode(1.6,5.3){Tx1c}\psellipse[linestyle=dashed](.5,5.3)(.2,.5)
\rput[l](.1,6){\small{$\xv^{[1]}_1$}}

\psframe(5.4,4.8)(6,5.8) \pscircle(5.7,5){.1} \pscircle(5.7,5.3){.1}
\pscircle(5.7,5.6){.1} \pnode(7,5.3){Rx1}
\pnode(6,5){Rx1a}\pnode(6,5.3){Rx1b}\pnode(6,5.6){Rx1c}\ncline[linestyle=dashed]{Rx1}{Rx1a}\ncline[linestyle=dashed]{Rx1}{Rx1b}\ncline[linestyle=dashed]{Rx1}{Rx1c}
\pnode(5.4,5.3){Rx1d} \psellipse[linestyle=dashed](6.5,5.3)(.2,.5)
\rput[l](6.5,6){\small{$\xu^{[1]}_1$}}

\psframe(1,3.2)(1.6,4.2) \pscircle(1.3,3.4){.1} \pscircle(1.3,4){.1}
\pnode(0,3.7){Tx2}
\pnode(1,3.4){Tx2a}\pnode(1,4){Tx2b}\ncline[linestyle=dashed]{Tx2}{Tx2a}\ncline[linestyle=dashed]{Tx2}{Tx2b}
\pnode(1.6,3.7){Tx2c}\psellipse[linestyle=dashed](.5,3.7)(.2,.5)
\rput[l](.1,4.4){\small{$\xv^{[2]}_1$}}

\psframe(5.4,3.2)(6,4.2) \pscircle(5.7,3.4){.1}
\pscircle(5.7,3.7){.1} \pscircle(5.7,4){.1} \pnode(7,3.7){Rx2}
\pnode(6,3.4){Rx2a}\pnode(6,3.7){Rx2b}\pnode(6,4){Rx2c}\ncline[linestyle=dashed]{Rx2}{Rx2a}\ncline[linestyle=dashed]{Rx2}{Rx2b}\ncline[linestyle=dashed]{Rx2}{Rx2c}
\pnode(5.4,3.7){Rx2d} \psellipse[linestyle=dashed](6.5,3.7)(.2,.5)
\rput[l](6.5,4.4){\small{$\xu^{[2]}_1$}}

\psframe(1,1.6)(1.6,2.6) \pscircle(1.3,1.8){.1}
\pscircle(1.3,2.4){.1} \pnode(0,2.1){Tx3}
\pnode(1,1.8){Tx3a}\pnode(1,2.4){Tx3b}\ncline[linestyle=dashed]{Tx3}{Tx3a}\ncline[linestyle=dashed]{Tx3}{Tx3b}
\pnode(1.6,2.1){Tx3c}\psellipse[linestyle=dashed](.5,2.1)(.2,.5)
\rput[l](.1,2.8){\small{$\xv^{[3]}_1$}}

\psframe(5.4,1.6)(6,2.6) \pscircle(5.7,1.8){.1}
\pscircle(5.7,2.1){.1} \pscircle(5.7,2.4){.1} \pnode(7,2.1){Rx3}
\pnode(6,1.8){Rx3a}\pnode(6,2.1){Rx3b}\pnode(6,2.4){Rx3c}\ncline[linestyle=dashed]{Rx3}{Rx3a}\ncline[linestyle=dashed]{Rx3}{Rx3b}\ncline[linestyle=dashed]{Rx3}{Rx3c}
\pnode(5.4,2.1){Rx3d} \psellipse[linestyle=dashed](6.5,2.1)(.2,.5)
\rput[l](6.5,2.8){\small{$\xu^{[3]}_1$}}

\psframe(1,0)(1.6,1) \pscircle(1.3,.2){.1} \pscircle(1.3,.8){.1}
\pnode(0,.5){Tx4}
\pnode(1,.2){Tx4a}\pnode(1,.8){Tx4b}\ncline[linestyle=dashed]{Tx4}{Tx4a}\ncline[linestyle=dashed]{Tx4}{Tx4b}
\pnode(1.6,.5){Tx4c}\psellipse[linestyle=dashed](.5,.5)(.2,.5)
\rput[l](.1,1.2){\small{$\xv^{[4]}_1$}}

\psframe(5.4,0)(6,1) \pscircle(5.7,.2){.1} \pscircle(5.7,.5){.1}
\pscircle(5.7,.8){.1} \pnode(7,.5){Rx4}
\pnode(6,.2){Rx4a}\pnode(6,.5){Rx4b}\pnode(6,.8){Rx4c}\ncline[linestyle=dashed]{Rx4}{Rx4a}\ncline[linestyle=dashed]{Rx4}{Rx4b}\ncline[linestyle=dashed]{Rx4}{Rx4c}
\pnode(5.4,.5){Rx4d} \psellipse[linestyle=dashed](6.5,.5)(.2,.5)
\rput[l](6.5,1.2){\small{$\xu^{[4]}_1$}}

\ncline{->}{Tx1c}{Rx1d}\ncline{->}{Tx1c}{Rx2d}\ncline{->}{Tx1c}{Rx3d}\ncline{->}{Tx1c}{Rx4d}
\ncline{->}{Tx2c}{Rx2d}\ncline{->}{Tx2c}{Rx1d}\ncline{->}{Tx2c}{Rx3d}\ncline{->}{Tx2c}{Rx4d}
\ncline{->}{Tx3c}{Rx3d}\ncline{->}{Tx3c}{Rx1d}\ncline{->}{Tx3c}{Rx2d}\ncline{->}{Tx3c}{Rx4d}
\ncline{->}{Tx4c}{Rx4d}\ncline{->}{Tx4c}{Rx1d}\ncline{->}{Tx4c}{Rx2d}\ncline{->}{Tx4c}{Rx3d}
\end{pspicture}
} \caption{Sample asymmetric and symmetric systems.}
 \label{fig:Figures}
\end{figure}
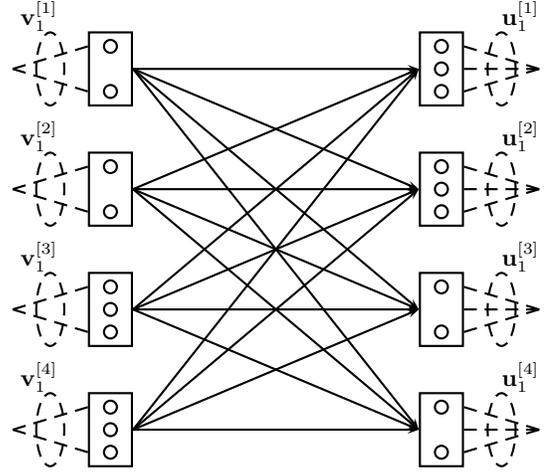
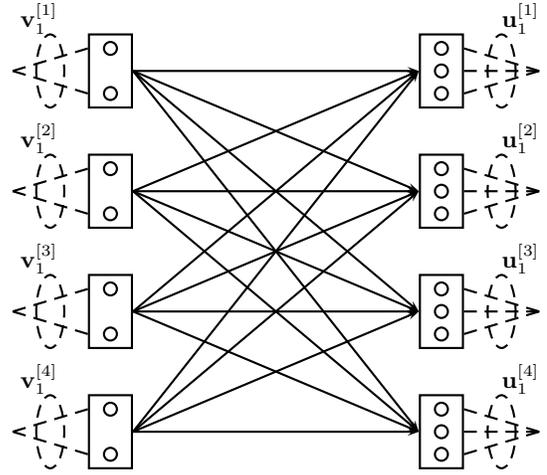

$\left(M^{[1]}\times N^{[1]}, d^{[1]}\right)\cdots(M^{[K]}\times
N^{[K]}, d^{[K]})$ denotes the $K$-user MIMO interference network,
where the $k^{th}$ transmitter and receiver have $M^{k}$ and $N^{k}$
antennas, respectively and the $k^{th}$ user demands $d^{[k]}$ DoF.
As defined earlier, $\left(M\times N, d\right)^K$ denotes the
$K$-user symmetric MIMO interference network, where each transmitter
and receiver has $M$ and $N$ antennas, respectively, and each user
demands $d$ DoF, so that the total DoF demand is $Kd$. Some sample
asymmetric and symmetric systems are shown in Fig.
\ref{fig:Figures}.

\section{Linear Interference Alignment Scheme}\label{sec:recalign}

In interference alignment precoding, the transmitted signal from the
$k^{th}$ user is $\xX^{[k]}=\xV^{[k]}\tilde{\xX}^{[k]}$, where
$\tilde{\xX}^{[k]}$ is a $d^{[k]}\times 1$ vector that denotes the
$d^{[k]}$ independently encoded streams transmitted from the
$k^{th}$ user. The $M^{[k]}\times d^{[k]}$ precoding filters
$\xV^{[k]}$ are designed to maximize the overlap of interference
signal subspaces at each receiver while ensuring that the desired
signal vectors at each receiver are linearly independent of the
interference subspace. Therefore, each receiver can zero-force all
the interference signals without zero-forcing any of the desired
signals. The zero-forcing filters at the receiver are denoted by
$\xU^{[k]}$. In \cite{55}, it is shown that an interference
alignment solution requires the simultaneous satisfiability of the
following conditions:
\begin{eqnarray}
\xU^{[k]\dagger}\xH^{[kj]}\xV^{[j]} &=& 0, \forall j\neq k \label{eqn:condition1}\\
\mbox{rank}\left(\xU^{[k]\dagger}\xH^{[kk]}\xV^{[k]}\right)&=&d^{[k]},
~~\forall k\in\{1,2,...,K\}, \label{eqn:condition2}
\end{eqnarray}
where $^\dagger$ denotes the conjugate transpose operator. Very importantly, \cite{55} explains how the condition (\ref{eqn:condition2}) is automatically satisfied almost surely if the channel matrices do not have any special structure, ${\mbox{rank}(\xU^{[k]})=\mbox{rank} (\xV^{[k]})=d^{[k]}\geq \min(M^{[k]},N^{[k]}})$ and $\xU^{[k]}, \xV^{[k]}$ are designed to satisfy (\ref{eqn:condition1}), which is independent of all direct channels ${\xH^{[kk]}}$.

The scenarios where it is difficult to theoretically determine the feasibility of interference alignment can be evaluated numerically using an iterative algorithm proposed in \cite{55}.   In this work, we develop analytical criteria to predict the feasibility of interference alignment. Our approach is to count the number of equations and variables in (\ref{eqn:condition1}). We assume generic MIMO channels with no structure and force the required ranks of the transmit and receive filters by design. Thus, (\ref{eqn:condition2}) is satisfied automatically and we only need to count the number of variables and equations for (\ref{eqn:condition1}).

\section{Proper System} \label{sec:Propernetworks}
The notion of a proper network is based on counting the number
of equations and the number of variables involved in the bilinear equations (\ref{eqn:condition1}).  While the formal definition appears later, put simply, the system $\Pi_{k=1}^K\left(M^{[k]}\times N^{[k]}, d^{[k]}\right)$ network is proper if and only if the cardinality
of every subset of the equations is less than or equal to the number
of variables involved in that subset of equations. Otherwise, the network is considered improper.
 The reason for this classification is the following intuition that forms the basis of our approach in this paper.

{\it Key Insight :} {\it The interference alignment problem is almost surely feasible for proper systems and almost surely infeasible for improper systems.}

Instead of a formal proof, we refer to Bezout's theorem, which states that a system of $m$ generic polynomial equations in $m$ variables will almost surely have as many common solutions as the product of the degrees of the polynomials. For our case, it suffices if there is one solution. The feasibility conditions in our case correspond to a system of polynomial equations, whose generic nature is due to the generic channel matrices. Due to the special bilinear form of the equations it needs to be shown that the polynomials are sufficiently generic. We are able to construct the rigorous proof for the $(2\times 3, 1)^4$ case which is omitted here due to lack of space. We expect generalizations to all cases should be possible, albeit cumbersome.

Next we explicitly account for all equations and variables. Let us start with the total number of equations $N_e$ and the total number of variables $N_v$.

\subsection{Counting the Number of Equations $N_e$ and Variables $N_v$}
\par To obtain $N_e$ and $N_v$, we rewrite
the condition in (\ref{eqn:condition1}) as follows:
\begin{equation}
\xu^{[k]\dagger}_m\xH^{[kj]}\xv^{[j]}_n = 0,~~ j\neq k, ~
j, k\in\{1,2,...,K\} \label{eqn:condition1b}
\end{equation}
\begin{equation*}
\forall n\in\{1,2,...,d^{[j]}\} \textrm{ and } \forall
m\in\{1,2,...,d^{[k]}\}
\end{equation*}
where $\xv^{[j]}_n$ and $\xu^{[k]}_m$ are the transmit and the
receive beamforming vectors (columns of precoding and interference
suppression filters, respectively).
\par $N_e$ is directly obtained from (\ref{eqn:condition1b}) as follows:
\begin{equation}
N_e=\underset{{\substack{ k,j\in\mathcal{K} \\
k\neq j}}}{{\sum }}d^{[k]}d^{[j]}. \label{eqn:Ne}
\end{equation}

However, calculating the number of variables $N_v$ is less straightforward. In particular, we have to be careful to not count any superfluous variables that do not help with interference alignment.

At the $k^{th}$
transmitter, the number of the $M^{[k]}\times 1$ transmit
beamforming vectors to be designed is $d^{[k]}$
$\left(\xv^{[k]}_n\textrm{, } \forall
n\in\{1,2,...,d^{[k]}\}\right)$. Therefore, at first sight it may seem that the precoding filter of
the $k^{th}$ transmitter, $\xV^{[k]}$, has $d^{[k]}M^{[k]}$
variables. However, as we argue next, without loss of generality one can eliminate $(d^{[k]})^2$ of these variables.

The $d^{[k]}$ linearly independent columns of the transmit precoding matrix ${\bf V}^{[k]}$  span the transmitted signal space
\begin{eqnarray}
\mathcal{T}^{[k]}&=&\mbox{span}({\bf V}^{[k]})\\
&=&\{{\bf v}: \exists {\bf a}\in\mathbb{C}^{d^{[k]}\times 1}, ~{\bf v}={\bf V}^{[k]}{\bf a} \}.
\end{eqnarray}
Thus, the columns of ${\bf V}^{[k]}$ are the basis for the transmitted signal space. However, the basis representation is not unique for a given subspace. In particular, consider any full rank $d^{[k]}\times d^{[k]}$ matrix ${\bf B}$. Then
\begin{eqnarray}
\mathcal{T}^{[k]}&=&\mbox{span}({\bf V}^{[k]})\\
&=&\{{\bf v}: \exists {\bf a}\in\mathbb{C}^{d^{[k]}\times 1}, ~{\bf v}={\bf V}^{[k]}{\bf a} \}\\
&=&\{{\bf v}: \exists {\bf a}\in\mathbb{C}^{d^{[k]}\times 1}, ~{\bf v}={\bf V}^{[k]}{\bf B}^{-1}{\bf Ba} \}\\
&=&\mbox{span}({\bf V}^{[k]}{\bf B}^{-1}).
\end{eqnarray}
Thus, post-multiplication of the transmit precoding matrix with any invertible matrix on the right does not change the transmitted signal subspace. Suppose we choose ${\bf B}$ to be the $d^{[k]}\times d^{[k]}$ matrix that is obtained by deleting the bottom $M^{[k]}-d^{[k]}$ rows of ${\bf V}^{[k]}$. Then, we have ${\bf V}^{[k]}{\bf B}^{-1}=\tilde{\xV}^{[k]}$, which is a $M^{[k]}\times d^{[k]}$ matrix with the following structure:
\begin{equation*}
\tilde{\xV}^{[k]}=\left[
\begin{array}{ccccc}
 &  & \xI_{d^{[k]}}  &  &\\
\bar{\xv}_1 & \bar{\xv}_2 & \bar{\xv}_3  & \cdots  & \bar{\xv}_{d^{[k]}}%
\end{array}%
\right]
\end{equation*}
where $\xI_{d^{[k]}}$ is the $d^{[k]}\times d^{[k]}$ identity matrix and $\bar{\xv}_n$,
$\forall n\in\{1,2,...,d^{[k]}\}$ are $\left(M^{[k]}-d^{[k]}\right)\times 1$ vectors.
It is easy to argue that there is no other basis representation for the transmitted signal space with fewer variables.

Therefore, by eliminating all superfluous variables for the interference alignment problem, the
number of variables to be designed for the precoding filter of
the $k^{th}$ transmitter, $\tilde{\xV}^{[k]}$, is
$d^{[k]}\left(M^{[k]}-d^{[k]}\right)$. Likewise, the actual number
of variables to be designed for the interference suppression
filter of the $k^{th}$ receiver, $\tilde{\xU}^{[k]}$, is
$d^{[k]}\left(N^{[k]}-d^{[k]}\right)$. As a result, the total number
of variables in the network to be designed is:
\begin{equation}
N_v=\sum_{k=1}^Kd^{[k]}\left(M^{[k]}+N^{[k]}-2d^{[k]}\right).\label{eqn:Nv}
\end{equation}

\subsection{Proper System Characterization}
 To formalize the definition of a proper system,  we introduce some notation.  We use the notation $E_{kj}^{mn}$ to represent the equation
\begin{eqnarray}
\xu^{[k]\dagger}_m\xH^{[kj]}\xv^{[j]}_n = 0.
\end{eqnarray}
The set of variables involved in an equation $E$ are indicated by the function $\mbox{var}(E)$. Clearly
\begin{eqnarray}
|\mbox{var}(E_{kj}^{mn})|=(M^{[j]}-d^{[j]})+(N^{[k]}-d^{[k]})
\end{eqnarray}
where $|\cdot|$ is the cardinality of a set.

Using this notation, we denote the set of $N_e$ equations as follows:
\begin{eqnarray*}
\mathcal{E}&=&\{E_{kj}^{mn}|~ j,k\in\mathcal{K}, k\neq j, \\
&&~~~~~~~~~m\in\{1,\cdots, d^{[k]}\}, n\in\{1,\cdots,d^{[j]}\}\}.
\end{eqnarray*}
This leads us to the formal definition of a proper system.
{\it Definition:} A $\Pi_{k=1}^K(M^{[k]}\times N^{[k]},d^{[k]})$ system is proper if and only if
\begin{eqnarray}
\forall S\subset\mathcal{E}, |S|\leq\left| \bigcup_{E\in
S}\mbox{var}(E)\right|. \label{eqn:proper}
\end{eqnarray}
In other words, for all subsets of equations, the number of variables involved must be at least as large as the number of equations in that subset.

Condition (\ref{eqn:proper}) provides us a way to predict the feasibility of interference alignment in a general $\Pi_{k=1}^K(M^{[k]}\times N^{[k]},d^{[k]})$ system. However, note that it can be computationally cumbersome because we have to test all subsets of the set of all equations. In most cases however, especially if the system is improper, simply comparing the total number of equations and the total number of variables may suffice.
\theorem
A $\Pi_{k=1}^K(M^{[k]}\times N^{[k]},d^{[k]})$ system is improper if $N_v<N_e$, i.e.,
\begin{eqnarray}
\sum_{k=1}^Kd^{[k]}\left(M^{[k]}+N^{[k]}-2d^{[k]}\right)<\sum_{{\substack{ k,j\in\mathcal{K} \\
k\neq j}}}^Kd^{[k]}d^{[j]}.\label{eq:generalcondition}
\end{eqnarray}

\example Consider the $(5\times 5,3)(5\times 5,2)^3$ system, i.e. a
$4$-user interference networks where all nodes have $5$ antennas,
user $1$ demands $3$ DoF and users $2,3,4$ demand $2$ DoF each.
There are a total of $60$ equations, so that there are $2^{60}-1$
subsets of the set of all equations. Testing each of them could be
very challenging. However, since the total number of variables
$N_v=48$ is less than the number of equations, the system is easily
seen to be improper.

Similarly, one can sometimes identify the bottleneck equations in the system by checking the equations with the fewest number of variables, i.e., the equations involving the transmitters and receivers with the fewest antennas.
\example
As a simple example, consider the system $(2\times 1, 1)^2$ which is clearly feasible (proper) because simple zero-forcing is enough for achievability. However, now consider the $(2\times 1, 1)(1\times 2,1)$ system, which also has the same total number of equations $N_e$ and variables $N_v$ as the $(2\times 1,1)^2$ system.  Thus, only comparing $N_v, N_e$ would lead one to believe that this system is proper. However, suppose we check $E_{12}^{11}$ which connects transmitter 2 and receiver 1, both of which have only one antenna each, i.e., let $S=\{E_{12}^{11}\}$ so that $|S|=1$. However $\mbox{var}(E_{12}^{11})=0$. Thus, this system has an equation with zero variables, which makes it improper, and therefore infeasible.

\example \label{ex:2x3,3x2}Several interesting cases emerge from
applying the condition (\ref{eq:generalcondition}). For example,
consider the $2$ user interference channel $(2\times 3, 1)(3\times
2, 1)$ where a total of $2$ degrees of freedom are desired. It is
easily checked that this system is proper and the achievable scheme
is described in \cite{93}. However, now consider the $4$-user
interference channel consisting of two sets of these channels, all
interfering with each other $(2\times 3,1)^2(3\times 2,1)^2$ which
is a $4$-user interference channel and a total of $4$ degrees of
freedom are desired. It is easily verified that this is a proper
system, i.e. it satisfies (\ref{eq:generalcondition}). It turns out
that a closed form solution for alignment can be found in this case.
It is somewhat surprising that going from two users to four users
simply doubled the degrees of freedom in this case, without any
penalty for interference between the users in terms of degrees of
freedom. Since in this paper our focus is only on feasibility and
not on closed form solutions, we omit the details of this example.

In general, testing if a large asymmetric system is proper can be
cumbersome. However, for the case of symmetric systems of the type
$(M\times N, d)^K$, the process is much simpler.
\subsection{Symmetric Systems $(M\times N,d)^K$}
\begin{theorem}\label{theorem:symmetric}
A symmetric system $(M\times N,d)^K$, is proper if and only if $N_v\geq N_e$. Equivalently, the system is proper iff
\begin{eqnarray}
M+N-(K+1)d\geq 0
\end{eqnarray}
\end{theorem}

{\it Remark:} Note that the condition $d\leq\min(M,N)$ is always
assumed even if it is not explicitly stated every time.

{\it Remark:} Theorem \ref{theorem:symmetric} implies that for every user to achieve $d$ degrees of freedom in a $K$ user interference channel, it suffices to have a total of $M+N\geq (K+1)d$ antennas between the transmitter and receiver of a user. The antennas can be distributed among the transmitter and receiver arbitrarily, as long as each of them has at least $d$ antennas. In particular, to achieve $K$ degrees of freedom in a $K$ user symmetric network we only need a total of $K+1$ antennas between the transmitter and receiver of each user. The system $(2\times 3,1)^4$ is such an example.

{\it Proof:} Because of the symmetry each equation is involved with the same number of variables and any deficiency in the number of variables shows up in the comparison of the total number of variables versus the total number of equations. Plugging in the values of $N_v, N_e$ computed earlier, we have the result of Theorem \ref{theorem:symmetric}.
\hfill\QED

\example
Consider $(1\times 2, 1)^3$, i.e. a 3-user MIMO symmetric interference
network, where each transmitter has one antenna, each receiver has two antennas, and each user demands $1$ DoF.  For this system $M+N-(K+1)d = 1+2-(4)<0$, so that this system is not proper.

\example \label{ex:2x3,1} Consider the $\left(2\times 3, 1\right)^4$
system. For this system, $M+N-(K+1)d=2+3-(5)=0$ which means this
system is proper.

\example \label{ex:5x5,2}Consider the $\left(5\times 5, 2\right)^4$
system. For this system $M+N-(K+1)d=5+5-10=0$, which means this
system is proper.

The following corollary shows the limitations of linear interference alignment over constant MIMO channels (with no symbol extensions).
\begin{corollary}
The ratio of the sum degrees of a proper  $(M\times N, d)^K$ system, normalized by a single user's degree of freedom in the absence of interference is bounded by:
\begin{eqnarray}
\frac{dK}{\min(M,N)}\leq 1+\frac{\max(M,N)}{\min(M,N)}-\frac{d}{\min(M,N)}
\end{eqnarray}
\end{corollary}
{\it Proof:} The proof is straightforward from the condition of Theorem \ref{theorem:symmetric}.\hfill\QED

{\it Remark: } For the case that $M=N$ note that the total number of degrees of freedom for a proper system is no more than twice the number of degrees of freedom achieved by each user in the absence of interference. Note that for diagonal (time-varying) channels it was shown in \cite{85} that the total number of degrees of freedom is $K/2$ times the number of degrees of freedom achieved by each user in the absence of interference. This result shows that the diagonal structure of the channel matrix is very helpful. Going from the case of no structure (arbitrary MIMO channels) to diagonal structure the ratio of total degrees of freedom to the single user degrees of freedom increases from a maximum value of $2$ to $K/2$.

The following corollary identifies the groups of symmetric systems,
which are either all proper or all improper.
\begin{corollary}
If the $(M\times N, d)^K$ system is proper (improper) then so is the $((M+1)\times(N-1),d)^K$ system, as long as
$d\leq\min(M,N-1)$.
Similarly, if  the $(M\times N, d)^K$ system is proper (improper) then so is the $((M-1)\times(N+1),d)^K$ system, as long as $d\leq\min(M-1,N)$.
\end{corollary}
{\it Proof:} Since the condition in Theorem \ref{theorem:symmetric} depends only on $M+N$, it is clear that one can switch transmit and receive antennas without affecting the proper (or improper) status of the system.

\example
 The systems $(1\times 4, 1)^4, (2\times 3,1)^4,(3\times 2,1)^4, (4\times 1,1)^4$ are in the same group, formed by switching between transmit and receive antennas. It is easy to see that the $(4\times 1,1)^4$ system is proper, because simple zero-forcing suffices to achieve the DoF demand. By virtue of being in the same group, the rest are proper as well.

\example \label{ex:2x8,2} Consider a $(2\times 8, 2)^ 4$ symmetric
system. Again, the DoF for this interference channel can be
trivially obtained by zero-forcing at each receiver. By switching
antennas from each transmitter to receiver, an equivalent $(5\times
5, 2)^ 4$ symmetric system of Example \ref{ex:5x5,2} is obtained,
which is also proper.

\section{Numerical Results}\label{sec:NumResults}
We tested numerous interference alignment problems, both symmetric
and asymmetric, and especially including each of the examples
presented in this paper, using the numerical algorithms of
\cite{55}. In every case so far, we have found the results to be
consistent with the guiding intuition of this work, proper systems
are feasible and improper systems are not.

\par In this section we provide numerical results for a few
interesting and representative cases. The results are in terms of
the leakage interference, as defined in \cite{55} - i.e. the
fraction of the interference power that is present in the dimensions
reserved for the desired signal. The interference percentage at the
$k^{th}$ receiver is evaluated as follows:
\begin{equation}\label{eqn:leakage}
p_{k}=\frac{\underset{j=1}{\overset{d^{[k]}}{\sum }}\lambda
_{j}\left[\mathbf{Q}^{[k]}\right] }{\textrm{Tr}[\mathbf{Q}^{[k]}]},
\end{equation}
where $\lambda_j$ denotes the smallest eigenvalue of a matrix, Tr
denotes the trace of a matrix and $\mathbf{Q}^{[k]}$ denotes the
interference covariance matrix at the $k^{th}$ receiver:
\begin{equation*}
\mathbf{Q}^{[k]}=\sum_{j=1,j\neq
k}^K\frac{P^{[j]}}{d^{[j]}}\xH^{[kj]}\xV^{[j]}\xV^{[j]\dagger}\xH^{[kj]\dagger}.
\end{equation*}
The numerator and the denominator of (\ref{eqn:leakage}) are the
interference and desired signal space powers at the $k^{th}$
receiver, respectively.
\par
\begin{figure} [htb]
\centering
\includegraphics[height=8cm, width=12cm] {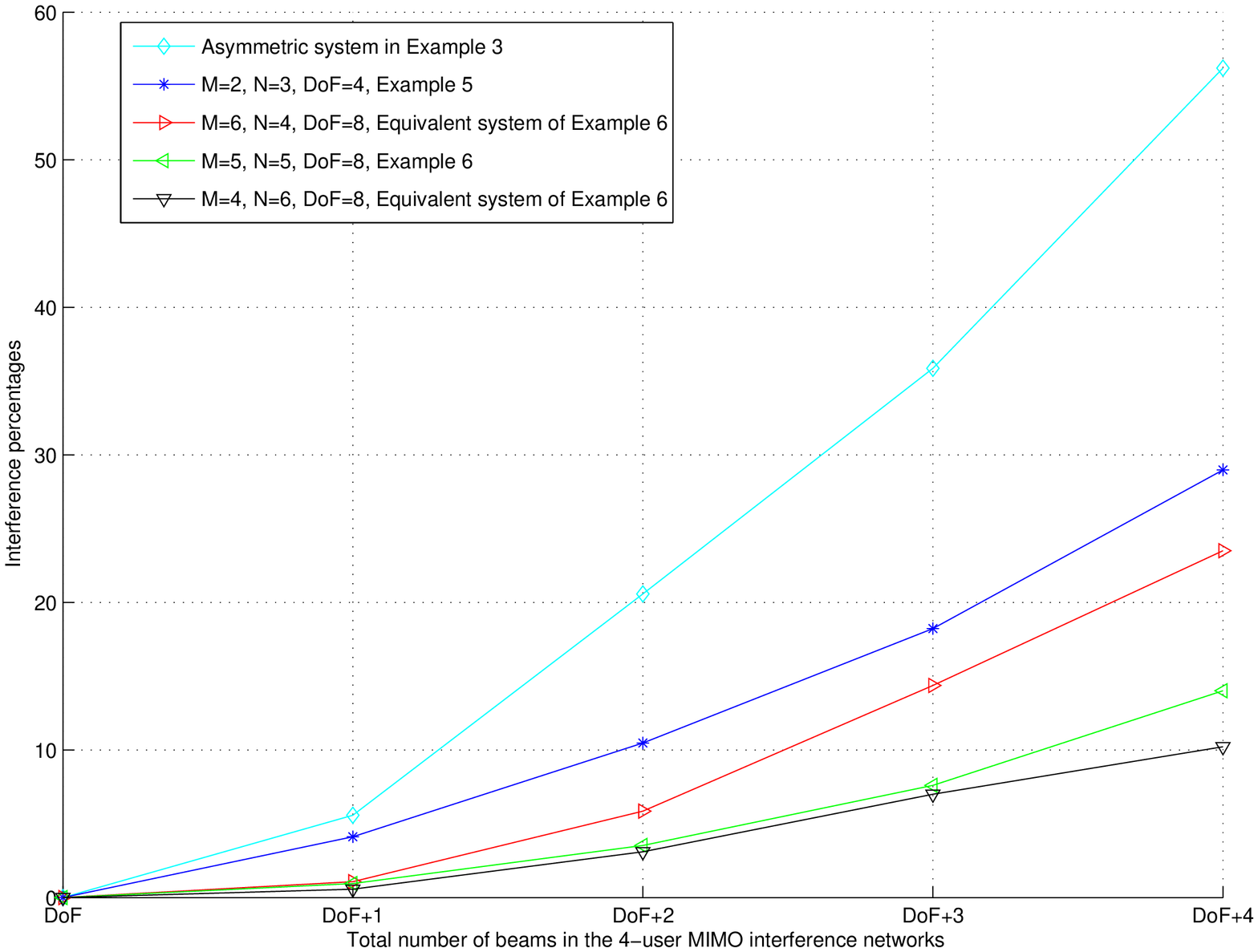}
\caption{Interference percentages as a function of the total number
of beams in the networks. \label{fig:DoFs}}
\end{figure}

 In Fig. \ref{fig:DoFs}, the interference percentages versus the total number
of beams are presented. The total number of beams are started from
the DoF of each network. Therefore, after the first point on the
x-axis, interference percentage of each network is not zero. The
non-zero interference percentage indicates that interference
alignment is not possible.
\section{Conclusion}\label{sec:Conclusion}
We propose an analytical method to predict the feasibility of interference alignment via linear schemes over constant MIMO channels that have no structure (i.e. no symbol extensions). Our approach to determine feasibility is to count the number of equations and variables. We define a system as proper if the number of variables is not smaller than the number of equations, and as improper otherwise, with the understanding that proper systems are feasible almost surely, while improper systems are almost surely infeasible. This observation follows from Bezout's theorem and the generic nature of the channel coefficients, and can be rigorously shown in some cases, to be included in the full paper. The full paper will also include closed form solutions for interference alignment for several cases, such as $(2\times 3, 1)^2(3\times 2,1)^2, (3\times 3,1)(2\times 3,1)^3, (2\times 4,1)(2\times 3)^3, (2\times 4,1)^2(4\times 2,1)^2(3\times 3,1)$ etc.

\section*{Acknowledgment}
The first author would like to thank V. R. Cadambe at University of
California, Irvine for valuable discussions on interference
alignment.
\bibliographystyle{IEEEtran}
\bibliography{IEEEabrv,IEEEfull}
\end{document}